\documentclass[iop]{emulateapj}
\usepackage{apjfonts}

\usepackage{booktabs}
\usepackage{color}
\usepackage{amsmath}

\begin{document}


\title{Turbulent Reconnection in Relativistic Plasmas And Effects of Compressibility}

\author{Makoto Takamoto}
\affil{Max-Planck-Institut f\"ur Kernphysik, Heidelberg, Germany}
\affil{Departmemt of Earth and Planetary Science, The University of Tokyo, Japan}
\email{mtakamoto@eps.s.u-tokyo.ac.jp}

\author{Tsuyoshi Inoue}
\affil{Division of Theoretical Astronomy, National Astronomical Observatory of Japan}
\email{tsuyoshi.inoue@nao.ac.jp}

\author{Alexandre Lazarian}
\affil{Department of Astronomy, University of Wisconsin, 475 North Charter Street, Madison, WI 53706, USA}
\email{alazarian@facstaff.wisc.edu}




\begin{abstract}
We report turbulence effects on magnetic reconnection in relativistic plasmas 
using 3-dimensional relativistic resistive magnetohydrodynamics simulations. 
We found 
reconnection rate became independent of the plasma resistivity due to turbulence effects similarly to non-relativistic cases. 
We also found 
compressible turbulence effects modified the turbulent reconnection rate predicted in non-relativistic incompressible plasmas; 
The reconnection rate saturates and even decays as the injected velocity approaches to the Alfv\'en velocity. 
Our results indicate  
the compressibility cannot be neglected when compressible component becomes about half of incompressible mode 
occurring when the Alfv\'en Mach number reaches about $0.3$. 
The obtained maximum reconnection rate is around $0.05$ to $0.1$, 
which will be able to reach around $0.1$ to $0.2$ if injection scales are comparable to the sheet length. 
\end{abstract}


\keywords{magnetic fields, magnetohydrodynamics (MHD), magnetic reconnection, relativistic processes, plasmas, turbulence}



\section{\label{sec:sec1}Introduction}

Magnetic reconnection is known as a process 
responsible for a very efficient magnetic field dissipation in many plasma phenomena. 
In particular, 
it is expected to play an important role for the acceleration of relativistic outflow in high energy astrophysical phenomena 
accompanying Poynting-dominated plasmas, 
such as relativistic jets~\citep{1977MNRAS.179..433B,1984RvMP...56..255B,2007MNRAS.380...51K,2015ApJ...803...30K}, 
pulsar wind~\citep{1984ApJ...283..694K,1984ApJ...283..710K,2001ApJ...547..437L,2003ApJ...591..366K}, and 
gamma-ray bursts (GRB)~\citep{2003astro.ph.12347L,2011ApJ...726...90Z}. 
However, the classical theory of magnetic reconnection~\citep{1958IAUS....6..123S,1957JGR....62..509P} 
predicts that magnetic reconnection becomes very slow in high magnetic Reynolds number plasmas ($R_{\rm m} \sim 10^{10}$), 
and fails to explain observed dissipation timescale in space and astrophysical phenomena. 
To solve this problem, a lot of efforts have gone into finding a \textit{fast-reconnection} process 
that does not depend on the value of resistivity. 
Using the equation of continuity, 
the reconnection rate can be expressed as 
\begin{equation}
  \label{eq:1.1}
  \frac{v_{\rm in}}{c_A} = \frac{\rho_{\rm s}}{\rho_{\rm in}}\frac{v_{\rm s}}{c_A} \frac{\delta}{L}
  ,
\end{equation}
where the subscript ``\textit{in}'' and ``\textit{s}'' indicate the inflow and outflow region, respectively, 
$v_{\rm in}, v_{\rm s}$ are the inflow and outflow velocity, respectively, $c_A$ is the Alfv\'en velocity, 
$\rho$ is the mass density, $\delta$ is the sheet thickness, 
and $L$ is the sheet length. 
This equation shows that 
fast reconnection processes can be obtained by increasing 
the density ratio: $\rho_{\rm s}/\rho_{\rm in}$, ~\citep{1982PhRvL..49..323B}, 
the outflow velocity: $v_{\rm s}/c_A$, 
and the aspect ratio of sheets: $\delta/L$
~\citep{1986PhFl...29.1520B,2001EPS...53..473S,2007PhPl...14j0703L,2009PhPl...16k2102B,2010PhRvL.105w5002U,2013ApJ...775...50T,2014ApJ...783L..21S}. 

Turbulence has been considered as a key process 
that can accelerate magnetic field annihilation~\citep{1985PhFl...28..303M,2011PhRvE..83e6405E,2012ApJ...755...76T,2013PhRvL.110y5001H}. 
In particular, many astrophysical objects are considered to be high Reynolds number plasma, 
and it is natural to assume those plasma are in a turbulent state 
\footnote{
In Poynting dominated plasmas, 
relatively strong turbulence may be able to be induced by various ways, e.g., 
Kruskal-Schwarzchild type instability \citep{2010ApJ...725L.234L} and Richtmyer-Meshkov type instability \citep{2012ApJ...760...43I} at a shock front,
which induces turbulence with velocity dispersion up to $\Delta v_{\rm turb} / a \lesssim 1.5 / \sqrt{\sigma}$~\citep{2012ApJ...755...76T}; 
the tearing instability with turbulent velocity Lorentz factor, $\gamma_{\rm turb} \simeq \sqrt{\alpha \sigma/2}$, 
where $\sigma$ is the magnetization parameter defined later, $a$ is the sound velocity, and $\alpha$ is energy conversion factor from magnetic field into kinetic energy. 
}.
It was theoretically suggested that 
strong Alfv\'enic turbulence also increases the sheet aspect ratio, and the reconnection rate becomes independent of the resistivity
(\citet{1999ApJ...517..700L}, henceforth LV99). 
LV99 predicts the following expression of reconnection rate: 
\begin{equation}
  \label{eq:1.2}
  \frac{v_{\rm in}}{c_A} \simeq \mathrm{min}\left[ \left( \frac{L}{l} \right)^{1/2}, \left( \frac{l}{L} \right)^{1/2} \right] \left( \frac{v_l}{c_A} \right)^2 
\end{equation}
where $l$ and $v_l$ are the energy injection scale and velocity dispersion of turbulence at the injection scale, respectively. 
This was examined using magnetohydrodynamics (MHD) simulation~\citep{2009ApJ...700...63K}. 
However, 
the numerical work  
was limited only in the non-relativistic incompressible 
regime 
with plasma $\beta$ larger than unity, 
and its applicability to relativistic Poynting dominated plasma with relativistic turbulence was unclear, 
which is very important in the context of high energy astrophysical phenomena~\citep{2013SSRv..178..459L,2015ApJ...802..113K}. 

In this paper, 
we extend the previous work to relativistic plasma including both matter and Poynting dominated plasma. 
We also investigate effects from compressibility on reconnection rate. 
In Section~2 we introduce the numerical setup and the method for the turbulence injection. 
The numerical result is presented in Section~3, 
and its theoretical explanation is presented in Section~4. 
Their implications for some high energy astrophysical phenomena are discussed in Section~5.
Section~6 summarizes our conclusions. 

\section{\label{sec:sec2}Simulation Setup}
We modelled the evolution of a current sheet in a turbulent flow using 3-dimensional resistive relativistic magnetohydrodynamics (RRMHD). 
The initial current sheet is modelled by 
the relativistic Harris sheet~\citep{1966PhFl....9..277H,2003ApJ...591..366K} 
whose magnetic field is expressed as 
\begin{align}
  {\bf B} &= B_0 \tanh [ z / \lambda] {\bf e}_x + B_G {\bf e}_y 
  \label{eq:2.0.1}
  ,
\end{align}
where $\lambda$ is the half-thickness of the initial sheet, 
and $B_0$ and $B_G$ are the reconnecting magnetic field and guide field component, respectively. 
The pressure inside of the sheet is assumed to satisfy the pressure balance, 
and the upstream pressure is determined by the magnetization parameter $\sigma \equiv B^2 / 4 \pi \rho h c^2$ 
where $h = 1 + (\Gamma / (\Gamma - 1))(p / \rho c^2)$ is the specific enthalpy of relativistic ideal gas with $\Gamma = 4 / 3$, 
and $p, \rho, c$ are the gas pressure, mass density, and the light velocity, respectively. 
The initial temperature is assumed uniform, $\Theta \equiv k_{\rm B} T / m c^2 = 1$,  
where $k_{\rm B}, m$ are the Boltzmann constant and particle rest mass, respectively. 

The evolution of the plasma is calculated using a 3-dimensional RRMHD scheme developed by \citet{2011ApJ...735..113T} 
which solves the full RRMHD equations in a conservative fashion using the constrained transport algorithm. 
This allows us to treat the mass density, momentum, energy, and divergence of magnetic field to be conserved within machine round-off error. 
The resistivity, $\eta$, was assumed to be constant, typically $\eta / L c = 10^{-4}$. 
We followed the similar simulation setup used in \citet{2009ApJ...700...63K}. 
The numerical box is assumed $[-L/2, L/2] \times [0, L] \times [-L, L]$ 
where $L = 20 \lambda$
\footnote{
Note that $\lambda$ is the \textit{initial} half-width of the sheet, and is a constant. 
}
. 
Note that the z-direction size of the numerical box is twice larger than x,y-direction to reduce the influence by turbulence on the reconnection inflow around z-boundaries. 
We divided the numerical box into the homogeneous numerical cells with size: $\Delta = L / 512$. 
The timestep size is set as: $\Delta t = 0.1 \Delta/c$. 
We set the periodic boundary condition in y-direction and free boundary condition x and z-direction. 

In our model, 
we drive turbulence using a similar method described by \citet{1999ApJ...524..169M}. 
We add a divergence-free 3-velocity field, $\delta {\vec v}$, 
and an electric field determined consistently to the injected velocity 
\footnote{First, we splitted the electric field as: ${\vec E} = {\vec E}_{\rm dissip} - {\vec v} \times {\vec B}$. 
Then, the ideal part is replaced by $\delta {\vec E} = - {\vec v}_{\rm new} \times {\vec B}$ 
where ${\vec v}_{\rm new}$ is obtained by the relativistic addition law of ${\vec v}$ and $\delta {\vec v}$.}
at time intervals $\Delta t_{\rm inj}$ in a box region located around the current sheet: $[-l_x, l_x] \times [0, L] \times [-l_z, l_z]$ 
where $l_x, l_z$ are a scale length that is sufficiently larger than the injected turbulence eddy scale; 
$\Delta t_{\rm inj}$ is chosen to be shorter than the eddy turnover time at the injection scale $l$ : $\Delta t_{\rm inj} = l / 4 \pi v_{\rm inj,0}$ 
where $v_{\rm inj,0} = 0.15 c$ is a typical injection velocity in this study. 
We note that the dynamics of the turbulence becomes insensitive to the injection time interval 
as long as the injection time interval is around the eddy turnover time at the injection scale. 
Following \citep{2011ApJ...734...77I} and \citep{2012ApJ...755...76T}, 
the velocity field is described as: $\gamma \delta v^i = \sum_{\vec{k}} P(k) \sin (\vec{k}\cdot\vec{x} + \phi_{\vec{k}}^i)$ 
where $\gamma$ is the Lorentz factor of the injected velocity, $i$ covers $\{x,y,z\},$ and $\phi_{\vec{k}}^i$ is a random phase. 
The one-dimensional power spectrum of the velocity field is assumed flat, 
$k^2 P(k) \propto k^0$. 
The initial-field perpendicular Fourier components $k_{\perp} = \sqrt{k_y^2 + k_z^2}$ are chosen in a shell extending from $k_l - \Delta k$ to $k_l + \Delta k$ 
where $k_l L/2\pi = 16, \Delta k L/2\pi = 3$. 
Note that this scale size is a little larger than the initial sheet scale, and can be well-resolved by our present resolution. 
The parallel-field wave number $k_{||} = k_x$ is determined by $k_{||} = k_{\perp} v_{\rm inj}/c_{\rm A}$ 
where $v_{\rm inj} = \sqrt{\langle \delta v^2 \rangle}$ is the root-mean-square velocity. 
Since the injected turbulent velocity does not immediately follow the critical balance condition~\citep{1995ApJ...438..763G}, 
the turbulence at the injected scale is weak~\citep{2000JPlPh..63..447G}, 
which transits into the strong turbulence around the sheet width scale because of the energy cascade~\citep{2008ApJ...672L..61P,2012PhRvL.109b5004V,2015arXiv150906601M}. 
The weak MHD turbulence cascades the wave energy only perpendicular to the magnetic field, 
the turbulence strength, $\chi \equiv \tau_{\rm A}/\tau_{\rm NL} \simeq k_{\perp} v_{\lambda} / k_{||} c_A$, gradually increases up to unity, 
which results in the strong turbulence in the sheet~\citep{2012PhRvL.109b5004V,2015arXiv150906601M}.
Note that the \textit{injected} velocity $v_{\rm inj}$ is different from the velocity \textit{at the injection scale} $v_l$, 
which was first pointed out in LV99. 
At the injection scale, 
the weak MHD turbulence theory gives the following the energy cascade rate: $\epsilon_{\rm inj} \sim v_l^2 / \tau_{\rm 
NL
} \sim v_l^4 l_{||}/l_{\perp}^2 c_A$ 
where $\tau_{\rm NL} \sim (l_{\perp}/v_l)^2 / (l_{||}/c_A)$ is the distortion time of Alfv\'en wave packets. 
Combining this to the injected power: $v_{\rm inj}^2 / \Delta t_{\rm inj}$, 
we obtain 
\begin{equation}
  \label{eq:2.1}
  v_{\rm inj} \gtrsim \sqrt{\frac{\Delta t_{\rm inj} l_{||}}{c_A l_{\perp}^2}} v_l^2 \propto v_l^2
  ,
\end{equation}
where the inequality resulted from the excitation of compression modes. 
In other words, 
$v_{\rm inj}$ is related to the injected power by some external force or free energy; 
on the other hand, 
$v_l$ is the velocity resulted from the energy cascade of the weak MHD turbulence. 
We observed this relation in our simulations, 
and assume this in the following (see also \citep{1999ApJ...517..700L,2009ApJ...700...63K}). 

\begin{figure}[t]
  \includegraphics[width=6cm,bb=0 0 633 568,clip]{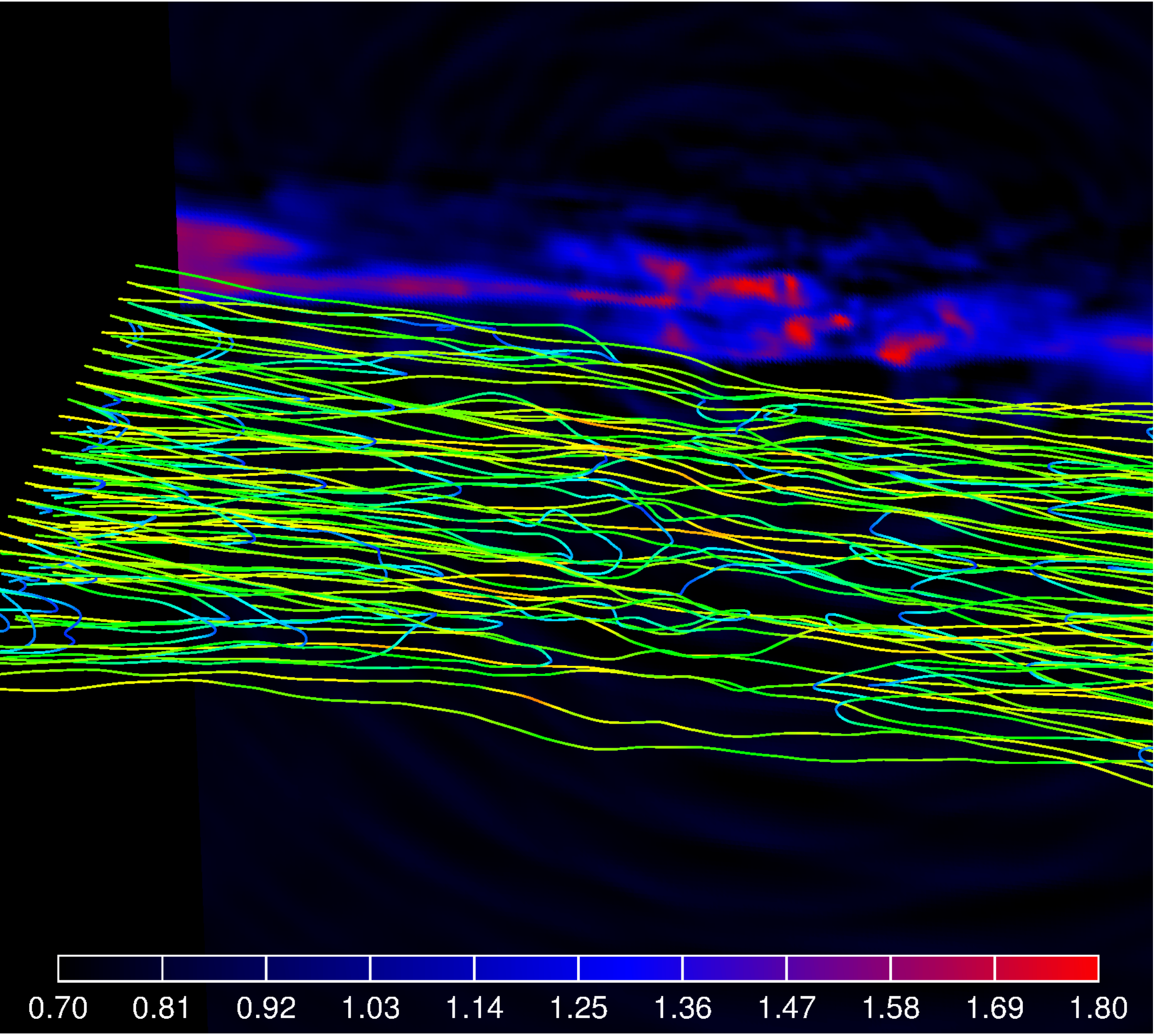}
  \label{fig:2.1}
  \caption{The profile of magnetic field lines in a turbulent sheet and the gas pressure (back plane)
           in the case of $\sigma = 5$. 
           The color bar is the gas pressure in the unit of the magnetic pressure in the initial inflow region. 
           }
\end{figure}

\section{\label{sec:sec3}Results}
Figure 1 is a snapshot of the gas pressure and magnetic field lines in the turbulent sheet in the case of highly magnetized case: $\sigma = 5$. 
Differently from the Sweet-Parker sheet, 
the sheet is highly stochastic due to the turbulence, 
which induces a lot of reconnection points in the sheet and drives a fast reconnection process. 
Figures 2 are the observed reconnection rates 
$v_{\rm R}$ 
which is measured using a method proposed by \citet{2009ApJ...700...63K} (see Equation (13) in this paper); 
this allows us to measure the effective value of $E_y/B_0$ in the 3-dimensional case, 
which provides us a reconnection inflow velocity less contaminated by turbulent flows than the direct measure of inflow velocity, $v_z$ 
\footnote{
In the following, the reconnection rate $v_{\rm R}$ obtained using the method by Kowal et al (2009) 
is identified by the reconnection inflow velocity $v_{\rm in}$, 
which becomes valid in a statistically steady state 
because $|E_y| \simeq v_{\rm in} B_0$ in this case. 
}
. 
The top panel shows the reconnection rate with respect to the injected turbulent velocity in various kinds of magnetized plasmas. 
This shows that the turbulent reconnection rate shows 3 characteristic behaviors depending on the injected turbulent velocity $v_{\rm inj}/c_A$: 
(1) increasing region following LV99; (2) saturation region giving maximum rate; (3) decreasing region.
When the injected turbulent velocity is sufficiently small, 
incompressible approximation can be applied, and the reconnection rate grows following Equation (\ref{eq:1.2}). 
On the other hand, 
when injected velocity becomes comparable to the Alfv\'en velocity, 
turbulence becomes compressible and the reconnection rate deviates from the incompressible theory. 
Interestingly, the injection velocity $v_{\rm inj}/c_A$ at the maximum rate becomes smaller as the magnetization parameter increases. 
We will discuss the relation of this tendency to the compressible effects in the next section. 
Note that the error bar in the panel seems decreasing with $\sigma$. 
We consider this is because the kinetic energy of turbulence becomes smaller comparing with the magnetic field energy 
as the magnetization parameter $\sigma$ increases. 
The bottom panel of Figure 2 is the reconnection rate with respect to the different Lundquist number. 
It shows the reconnection rate is independent of the Lundquist number, and determined by the turbulent strength. 
Note that the obtained maximum reconnection rate is very fast, 
$v_{\rm R}/c_A \sim 0.05$, 
and even comparable to the relativistic Petschek reconnection rate~\citep{2005MNRAS.358..113L}. 
This maximum reconnection rate also indicate that it will be possible to reach around $0.1$ to $0.2$ 
if injection scales are comparable to the sheet length as indicated by Equation (2)
\footnote{
Unfortunately, the dependence on the injection scale $l$ is hard to test 
because of the limited inertial range of turbulence in the simulation. 
}
. 

\begin{figure}[t]
  \includegraphics[width=7cm,bb=0 0 216 151,clip]{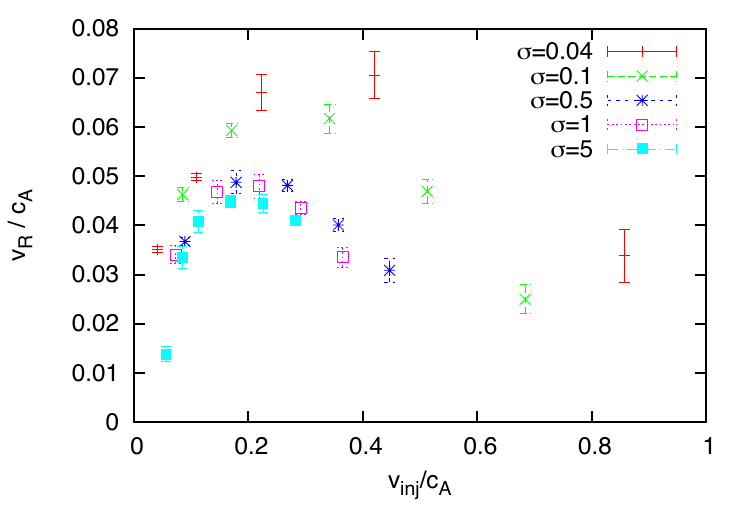}
  \includegraphics[width=7cm,bb=0 0 216 151,clip]{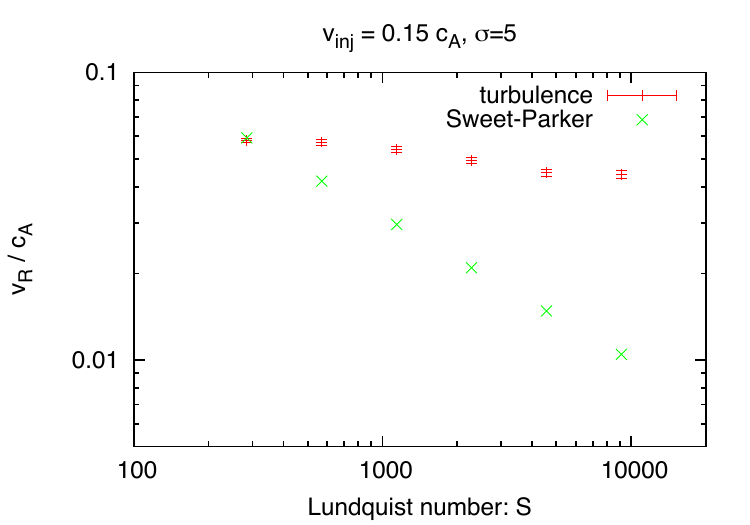}
  \label{fig:2.2}
  \caption{Observed reconnection rate in its steady state. 
           Top: Reconnection rate with respect to the injected turbulent velocity. 
           Bottom: Reconnection rate with respect to the Lundquist number: $S \equiv L c_A/\eta$. 
           }
\end{figure}

Figure 3 is the guide field dependence of reconnection rate in the case of $\Delta v_{\rm inj} = 0.15 c_A$ and $\sigma_{\rm R} = 5$ 
where $\sigma_R$ is the magnetization parameter determined by reconnecting magnetic field component $B_0$. 
We fixed the reconnected magnetic field and added the guide field component. 
As was reported by \citet{2009ApJ...700...63K}, 
the reconnection rate becomes independent of the guide field strength even in a relativistic Poynting-dominated plasma. 
In the case of $B_G/B_0 = 1$, the time scale necessary for reaching the steady state becomes 5 times longer than no guide field case. 

\begin{figure}[t]
  \includegraphics[width=7cm,bb=0 0 216 151,clip]{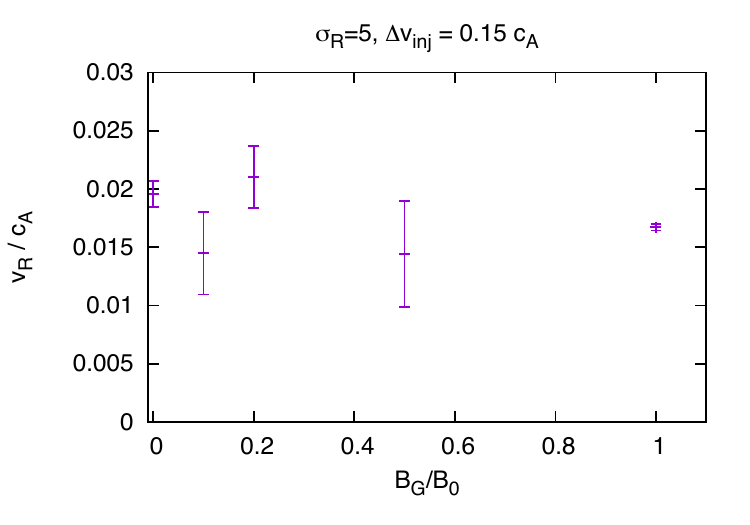}
  \label{fig:2.3}
  \caption{Observed reconnection rate in its steady state with respect to the guide field. 
           Note that the guide field is added to the fixed reconnection field, $B_x$, 
           so that the total magnetization parameter increases with the guide field. 
           }
\end{figure}


\section{\label{sec:sec4}Theoretical Considerations}
\subsection{\label{sec:sec4.1}Sheet Density}

\begin{figure}[t]
  \includegraphics[width=7cm,clip,bb=0 0 216 151]{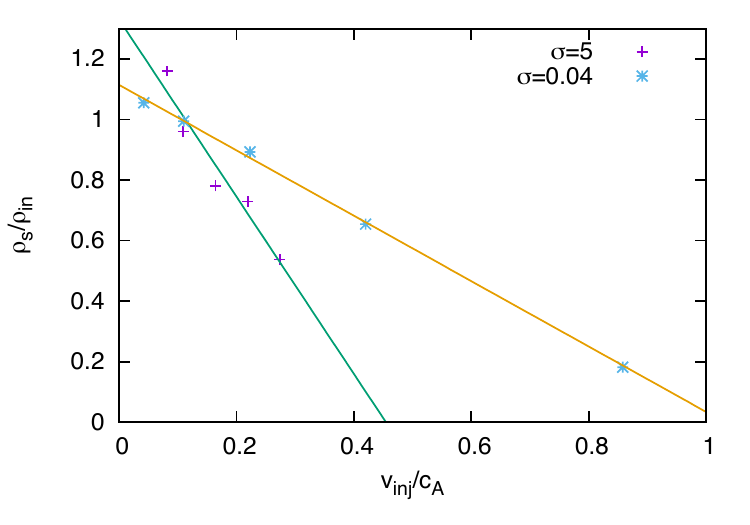}
  \label{fig:4.1}
\caption{The density ratio between that of inflow and sheet region: $\rho_{\rm s} / \rho_{\rm in}$ 
         in the cases of Poynting dominated case: $\sigma = 5$ and matter dominated case: $\sigma=0.04$. 
         The ratio decreases with increasing the turbulent strength due to the compressible effect.
         }
\end{figure}

The obtained reconnection rate in Figure 2 shows an interesting behavior owing to compressibility 
which cannot be explained by incompressible theory, Equation (\ref{eq:1.2}). 
In the following, 
we give an explanation for the saturation and depression of the reconnection rate in high Alfv\'en Mach number regime. 
Equation (\ref{eq:1.1}) indicates the compressible effects can be divided into 2 parts: 
(1) the density ratio between sheet and inflow region $\rho_{\rm s} / \rho_{\rm in}$; 
(2) decrease of the sheet width $\delta/L$. 
Note that $\delta$ is the actual sheet thickness determined by the turbulence which is different from the initial thickness $\lambda$. 

We begin with discussing the density ratio. 
Figure 4 plots $\rho_{\rm s} / \rho_{\rm in}$ with respect to the injected turbulence velocity 
in the matter and Poynting dominated cases $\sigma = 0.04$ and $5$, respectively. 
They show that the density ratio decreases linearly with the turbulent strength. 
This can be understood from the conservation of energy flux: 
\begin{align}
  \rho_{\rm in} h_{\rm in} c^2 \left(1 + \sigma \right) v_{\rm in} L &+ \rho_{\rm in} (1 + 2 h_{\rm in} \sigma) 
  \epsilon_{\rm inj} l_x l_z
  \nonumber
  \\
  &= \left(\rho_{\rm s} h_{\rm s} c^2 \gamma_{\rm s}^2 + \frac{B_{\rm s}^2}{4 \pi} \right) v_{\rm s} \delta
  \label{eq:4.1}
  . 
\end{align}
We assumed a non-relativistic inflow, $\gamma_{\rm in} = 1$. 
The 1st and 3rd terms are the energy flux in the inflow and outflow region, respectively. 
Note that the 2nd term in left-hand side of the equation expresses kinetic and electric field energy of the injected turbulence; 
The turbulent components in the sheet is neglected because we use a sub-Alfv\'enic turbulence
whose kinetic energy is small compared with the other terms. 
Using the pressure balance: $p_{\rm s} = p_{\rm in} + B_{\rm in}^2/8 \pi \gamma_{\rm in}^2$, 
the steady state condition: $c E_y = B_{\rm in} v_{\rm in} = B_{\rm s} v_{\rm s}$, 
and the equation of continuity, Equation (\ref{eq:1.1}), 
we obtain 
\begin{equation}
  \label{eq:4.3}
  \frac{\rho_{\rm s}}{\rho_{\rm in}} = \frac{(2 h_{\rm in} \sigma + 4 \Theta) \gamma_{\rm s}}{h_{\rm in} (1 + \sigma) - \gamma_{\rm s}} \left[1
   - \frac{1 + 2 h_{\rm in} \sigma}{(2 h_{\rm in} \sigma + 4 \Theta) \gamma_{\rm s}^2 } \frac{\epsilon_{\rm inj}}{\delta} 
    \frac{l_x l_z}{v_{\rm s} c^2} \right]
  , 
\end{equation}
where we neglected a small term proportional to $(\delta / L)^2 \ll 1$ resulting from $B_{\rm s}^2/4\pi$.  
Note that the denominator, $h_{\rm in} (1 + \sigma) - \gamma_{\rm s}$, is always positive because $\gamma_{\rm s} \lesssim \gamma_A = \sqrt{1 + \sigma}$ 
where $\gamma_A$ is the Lorentz factor of the Alfv\'en velocity in the upstream region. 
Since LV99 predicts $\delta \propto L (v_{\rm inj}/c_A)$ (see Equations (2) and (4)), 
the second term in Equation (\ref{eq:4.3}) becomes proportional to: $\epsilon_{\rm inj}/L v_{\rm inj} \propto v_{\rm inj}$; 
And we finally obtain the following form of relation: $\rho_{\rm s}/\rho_{\rm in} = \alpha (1 - \beta v_{\rm inj}/c_A)$ 
which qualitatively reproduces the linear dependence of the density ratio on the injected turbulence strength $v_{\rm inj}$ in Figure 4
~\footnote{
Our simulation results did not reproduce the exact value of $\alpha$ and $\beta$ indicated by Equation (\ref{eq:4.3}). 
However, the obtained values of $\alpha$ roughly reproduced the predicted value by Equation (\ref{eq:4.3}), 
and our results also reproduced the increasing nature of $\alpha$ and $\beta$ in terms of $\sigma$ parameter; 
this is indicated by Equation (\ref{eq:4.3}) with an assumption $v_s \sim c_A$ if $v_s < 0.3 c$ otherwise $v_s = 0.3 c$ 
which is a known result for relativistic Sweet-Parker sheet~\citep{2011ApJ...739L..53T}. 
Concerning $\beta$, note that it is difficult to estimate the exact value from simulation results because of the uncertainty of $l_x, l_z, \delta$. 
}.

This decrease of the sheet density can be explained as follows. 
When the turbulence energy injection rate is small, $v_{\rm inj} \ll c_A$, 
it increases the sheet width as predicted by LV99. 
However, the increase of the sheet width is proportional to $|v_{\rm inj}| \propto \epsilon_{\rm inj}^{1/2}$, 
and the turbulence energy injection rate $\epsilon_{\rm inj}$ cannot be absorbed into the sheet width expansion 
as indicated by the second term in Equation (\ref{eq:4.1}). 
In this case, in order to keep the energy flux conservation, the system reduces the inflow velocity $v_{\rm in}$ in the left-hand side of Equation (\ref{eq:4.1}), 
and this results in the decrease of the sheet density comparing with the inflow density as indicated in Equation (\ref{eq:1.1}). 
In other words, 
this is because the second term in the left-hand side increases with $\epsilon_{\rm inj}$ but right-hand side only increases with $\delta \propto \epsilon_{\rm inj}^{1/2}$. 
Hence, to keep the energy flux conservation, the system reduces the inflow velocity $v_{\rm in}$ in the left-hand side of Equation (\ref{eq:4.1}), 
and this results in the decrease of the sheet density comparing with the inflow density as indicated in Equation (\ref{eq:1.1}). 

Note that Equation (\ref{eq:4.3}) indicates the density in sheets becomes negative when a too strong turbulence is injected, 
which does not occur in real situations. 
This is prohibited by including neglected terms in Equation (\ref{eq:4.1}). 
In particular, as is discussed in Section \ref{sec:sec4.2}, energy flux escaping as compressible waves cannot be neglected 
as the injected turbulent Alfv\'en Mach number approaches unity. 

\subsection{\label{sec:sec4.2}Compressible Turbulence Effects}

\begin{figure}[t]
  \includegraphics[width=8.8cm,clip,bb=0 0 359 235]{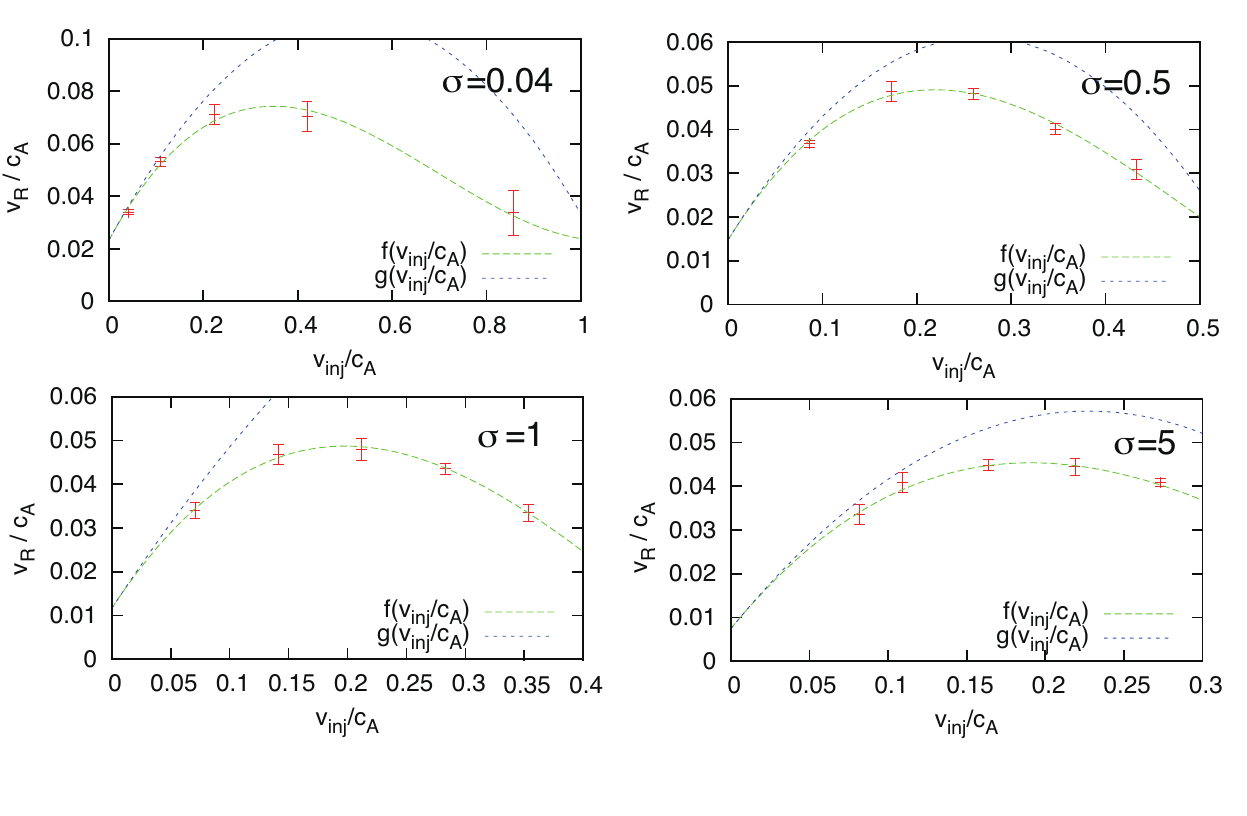}
  \label{fig:4.2}
  \caption{Reconnection rates fitted by two functions: $f(v_{\rm inj}/c_A) = C_1 (\rho_{\rm s}/\rho_{\rm in}) [v_{\rm inj}/c_A 
          - C_2 (v_{\rm inj}/c_A)^2]$, 
          $g(v_{\rm inj}/c_A) = C_1 (\rho_{\rm s}/\rho_{\rm in}) v_{\rm inj}/c_A$. 
          The function $f$ can explain the reconnection rate from the matter dominated case to the Poynting dominated case, 
          which indicates the compressible effects definitely affect the reconnection rate in turbulence. 
           }
\end{figure}

Next, 
LV99 obtained the following relation: $\delta/L \propto (v_l / c_A)^2 \propto v_{\rm inj}/c_A$ using the incompressible MHD turbulence cascade law. 
In this paper, we treated compressible MHD turbulence, 
so that it is expected the above relation should be modified. 
More precisely, the LV99's relation can be rewritten as: 
\begin{equation}
  \label{eq:4.3.0}
  \frac{\delta}{L} \sim \left( \frac{2 \epsilon_{\rm inj} l}{c_A^3} \right)^{1/2} {\rm min} \left[ \left(\frac{L}{l}\right)^{1/2}, \left(\frac{l}{L} \right)^{1/2} \right]
  ,
\end{equation}
and substituting, $\epsilon_{\rm inj} \sim v_l^4/2 l c_A$, 
recovers Equation (\ref{eq:1.2}). 
Hence, if we find an expression of the energy injection rate $\epsilon_{\rm inj}$ including compressible effects, 
Equation (\ref{eq:4.3.0}) may give us a new expression of the sheet width. 
Recently, \citet{2013PhRvE..87a3019B} obtained an exact relation of energy cascade rate 
in the non-relativistic isothermal MHD turbulence. 
In the strong background average magnetic field limit, 
the relation reduces to: 
\begin{equation}
  \label{eq:4.3.1}
  - 4 \epsilon = \nabla \cdot {\vec F}
  + B_0^2 S
\end{equation}
where the divergence $\nabla$ is performed on the correlation length which plays a role of the eddy scale length, 
${\vec F}$ is the energy flux vector including compressible effects with order of $B_0^2$, 
and $S$ is a source or sink term due to the compressible effects. 
This indicates that
the compressible effects cannot be neglected in the strong background magnetic field, 
and the energy cascade rate should be redefined as an effective mean total energy cascade rate: 
$\epsilon_{\rm eff} \equiv \epsilon + B^2_0 S/4$, 
and this will give us the necessary correction term in Equation (\ref{eq:4.3.0}). 
\footnote{
The source/sink term $S$ from compressible effects includes (1) mode exchange between the compressible modes and Alfv\'en mode; 
(2) a direct cascade of eddy size by dilatation ($\nabla \cdot {\vec v} > 0$) or compression ($\nabla \cdot {\vec v} < 0$) 
(more detailed discussion can be found in \citep{2013PhRvE..87a3019B}). 
It is our future work to determine which effect works dominantly. 
}
Performing the Taylor expansion of $\epsilon_{\rm eff}$ in $v_{\rm inj}/c_A < 1$ up to 2nd-order, 
the corrected sheet width can be written as: 
\begin{equation}
  \label{eq:4.4}
  \frac{\delta}{L} \simeq \mathrm{min}\left[ \left( \frac{L}{l} \right)^{1/2}, \left( \frac{l}{L} \right)^{1/2} \right] 
  \left[ \frac{v_{\rm inj}}{c_A} - C_2 \left( \frac{v_{\rm inj}}{c_A} \right)^2 \right]
  ,
\end{equation}
where $C_2$ is a coefficient resulting from the expansion. 
Figure 5 is the reconnection rates with various kinds of magnetization parameters: $\sigma = 0.04, 0.5, 1, 5$ 
which are fitted by 2 functions; 
one uses Equation (\ref{eq:4.4}) with the density ratio, Equation (\ref{eq:4.3}): 
$f(v_{\rm inj}/c_A) = C_1 (\rho_{\rm s}/\rho_{\rm in}) [v_{\rm inj}/c_A - C_2 (v_{\rm inj}/c_A)^2]$, 
and the other only takes into account the density ratio and uses LV99 sheet width, Equation (\ref{eq:4.4}): 
$g(v_{\rm inj}/c_A) = C_1 (\rho_{\rm s}/\rho_{\rm in}) v_{\rm inj}/c_A$. 
Note that $C_1$ and $C_2$ describe coefficients independent of the injection velocity $v_{\rm inj}$ indicated by Equation (\ref{eq:4.4}). 
As can be seen, they are well-reproduced only by $f$ whose $C_2$ are around unity in all the cases. 

\begin{figure}[t]
 \centering
  \includegraphics[width=7cm,clip,bb=0 0 216 151]{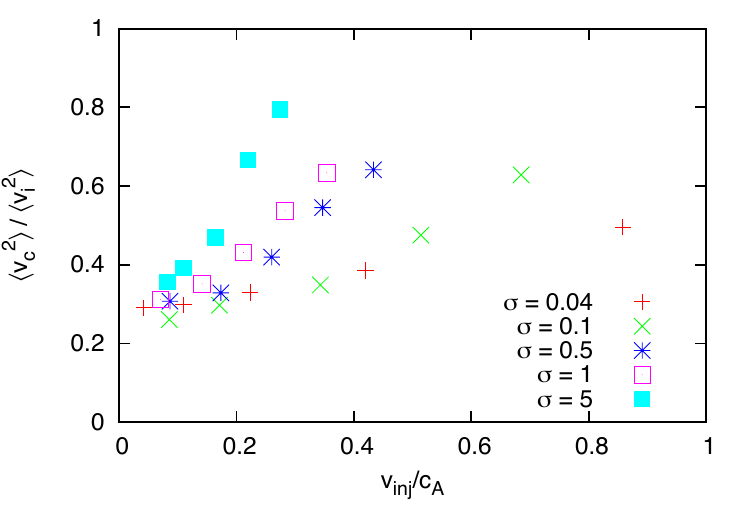}
  \includegraphics[width=7cm,clip,bb=0 0 216 151]{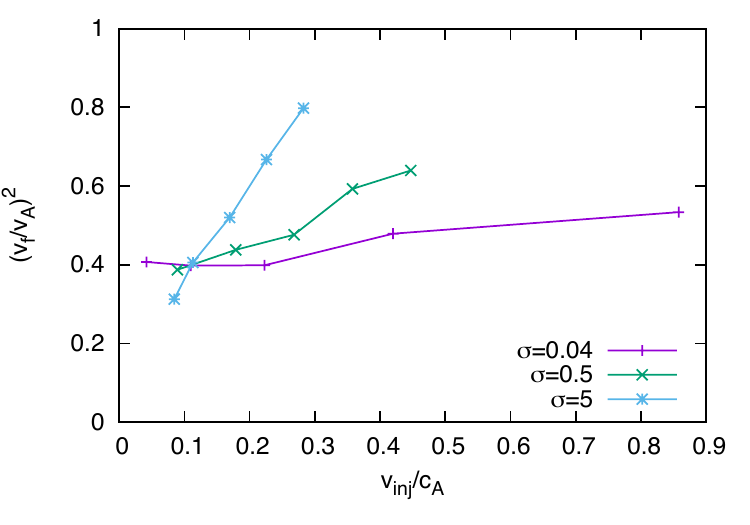}
  \caption{Top: The ratio of the compressible and incompressible velocity components: $\langle v_{\rm c}^2 \rangle$ and $\langle v_{\rm i}^2 \rangle$, respectively. 
           Bottom: The ratio of the fast MHD wave and Alfv\'en wave power: $(V_f/V_A)^2$. 
           The data were calculated by the background without sheet to avoid inhomogeneous back ground and back reaction from the sheet. 
           The panel shows the ratio with respect to the strength of the injected turbulence: $v_{\rm inj}/c_A$ with various kinds of the magnetization parameter. 
           }
  \label{fig:4.3}
\end{figure}

The above discussion assumes that the turbulence becomes compressible. 
To conform the validity of this assumption, 
we performed the Helmholtz decomposition and the MHD wave mode decomposition of the velocity field. 
The analysis were performed using a background obtained by new runs without current sheets but using the same setup. 
This is because the current sheets introduce a inhomogeneous background 
which makes it very difficult to perform the above decomposition. 
The top panel of Figure 6 are the ratio between the compressible and incompressible velocity components: 
$\langle v_{\rm c}^2 \rangle$ and $\langle v_{\rm i}^2 \rangle$, respectively 
in terms of the Alfv\'en Mach number of the injected turbulence velocity dispersion, $v_{\rm inj} / c_A$, 
using various magnetization parameter $\sigma$. 
As expected, this shows that in all cases the compressible component increases with the turbulent strength. 
We also note that the maximum reconnection rate in Figure 2 is obtained when  $\langle v_{\rm c}^2 \rangle / \langle v_{\rm i}^2 \rangle \sim 0.4$. 
Interestingly, Figure 6 shows the compressible component increases with the magnetization of the plasma 
similarly to in non-relativistic MHD turbulence~\citep{2002PhRvL..88x5001C}. 
We consider this is due to the $B^2_0$ factor in Equation (\ref{eq:4.3.1}), 
which indicates the compressible effects becomes more important as the background magnetic field increases. 
Unfortunately, Equation (\ref{eq:4.3.1}) is a result in the case of the non-relativistic MHD turbulence, 
and its relativistic extension is our future work. 
The bottom panel of Figure 6 is the ratio of the fast MHD wave to the Alfv\'en wave power. 
The wave decomposition was performed in the Fourier space assuming the linear dispersion relation 
similarly to the non-relativistic case by \citet{2002PhRvL..88x5001C}. 
It shows a very similar behavior obtained by the Helmholtz decomposition. 
Interestingly, we also found the slope of the ratio, $(V_f/V_A)^2$, is proportional to $\sqrt{\sigma}$, 
so that it can be written as: $(V_f/V_A)^2 \propto \sqrt{\sigma} v_{\rm inj}/c_A$. 
Equation (\ref{eq:4.4}) can effectively be derived by considering the wave decomposition. 
The turbulent reconnection theory in LV99 considers the MHD turbulence results in a wider current sheet 
because of the wandering motion of the magnetic field driven by Alfv\'en waves. 
Hence, $\epsilon_{\rm inj}$ in Equation (\ref{eq:4.3.0}) is equivalent to the Alfv\'en wave power, $V_A^2$, 
where $V_A$ is the Alfv\'en wave component of the velocity. 
In the compressible regime, a part of injected energy is distributed into the fast wave, 
and $\epsilon_{\rm inj}$ in Equation (\ref{eq:4.3.0}) should be rewritten as: 
\begin{equation}
  \label{eq:4.5}
  V_A^2 \simeq v_{\rm inj}^2 - V_f^2 \sim v_{\rm inj}^2 - V_A^2 \sqrt{\sigma} v_{\rm inj}/c_A
  ,
\end{equation}
where $V_f$ is the fast wave component of the velocity. 
This reduces to 
\begin{equation}
  \label{eq:4.6}
  \sqrt{V_A^2} \sim v_{\rm inj} \left[1 - \frac{\sqrt{\sigma}}{2} \frac{v_{\rm inj}}{c_A} \right]
  ,
\end{equation}
where we assume $v_{\rm inj}/c_A < 1$. 
This reproduces the dependence of Equation (\ref{eq:4.6}) on $v_{\rm inj}$
\footnote{
Note that slow waves basically propagates along magnetic fields, 
and they are also responsible for the magnetic field wandering. 
}. 
However, we cannot find the dependence of $C_2$ on $\sqrt{\sigma}$ indicated by Equation (\ref{eq:4.6}). 
We consider this may be due to the effect of inhomogeneous background structure and the back reaction from the tearing instability 
which are not taken into account to obtain Figure 6. 

\section{\label{sec:sec5}Applications to High Energy Astrophysical Phenomena}

In this section, we discuss applications to high energy astrophysical phenomena, 
that is, the Crab pulsar wind nebula, relativistic jets, and gamma-ray bursts. 
We estimate the necessary spatial and temporal scales for explaining those phenomena, 
and compare them with the actual observational indications. 
In those phenomena, it is natural to consider the driving source of turbulence depends on phenomena and the resulting turbulence strength is different. 
However, there are still a lot of theoretical and observational uncertainties in those phenomena, 
such as the precise spatial distribution of the magnetic field strength, the particle number density, and the particle composition 
which is necessary to estimate synchrotron radiation flux. 
And identifying each turbulent process is beyond the scope of this paper. 
Hence, in the following, we only use the values of reconnection rate to estimate those scales, 
so that the discussions can in general be applied for another dissipation mechanism, 
such as collisionless reconnection or plasmoid-chains. 
\\\\
\textit{--Crab Pulsar Wind Nebula}\\
In the case of the Crab pulsar wind nebula, 
the wind region is filled with current sheets (\textit{striped wind}). 
It is known that the magnetic field cannot be completely dissipated in the wind region~\citep{2001ApJ...547..437L,2003ApJ...591..366K}. 
One way to avoid this problem is to assume that the magnetic field dissipates just behind the termination shock
~\citep{2003MNRAS.345..153L,2007AA...473..683P,2011ApJ...741...39S}. 
If assuming high-$\sigma$ upstream flow, the down stream of the termination shock is still relativistic flow with Lorentz factor $\sqrt{\sigma}$. 
In the downstream rest frame, 
the sheet separation is around $\pi \sqrt{\sigma} r_{\rm LC}$ where $r_{\rm LC} = c / \Omega$ is the light cylinder radius
, 
and $\Omega$ is the rotation period of the Crab pulsar. 
The necessary timescale to dissipate the magnetic field between the sheets by reconnection with reconnection rate $v_{\rm in}$ is $\pi \sqrt{\sigma} r_{\rm LC} / v_{\rm in}$. 
In the termination shock rest frame, which would be equivalent to the observer's frame, 
additional Lorentz factor $\sqrt{\sigma}$ is multiplied to the dissipation time scale, and it becomes
\begin{equation}
  \label{eq:5.1}
  \tau_{\rm dissip,PWN} = \pi \sigma r_{\rm LC} / v_{\rm in}
  .
\end{equation}
During this timescale, the current sheets propagates 
\begin{align}
  \label{eq:5.2}
  l_{\rm dissip} &= c \tau_{\rm dissip,PWN} = \pi \sigma r_{\rm LC} c / v_{\rm in} 
  \nonumber
  \\
  &= 5 \times 10^{-5} [\mathrm{pc}] \left( \frac{2 \pi r_{\rm LC}}{10^9 [\mathrm{cm}]} \right) \left( \frac{\sigma}{10^4} \right) \left( \frac{v_{\rm in}/c}{0.1} \right)^{-1} 
  ,
\end{align}
which is sufficiently short compared to the Crab pulsar wind nebula scale size 
($\sim 1$ [pc]), 
and the turbulent reconnection can be one of the possible dissipation mechanisms to solve the $\sigma$-problem. 
The indicated value in Equation (\ref{eq:5.2}) is still too small to be resolved by X-ray (e.g. by the Chandra telescope). 
However, it may be possible to be observed by a future mission if $\sigma$ is larger than $10^4$ and the reconnection rate $v_{\rm in}/c$ is smaller than $0.1$. 
\\\\
\textit{--Relativistic Jets}\\
In this case, although there is no proof of the existence of current sheets in the observed jets, 
we assume a dynamo-process such as the magneto-rotational instability (MRI) in their accretion disk results in current sheets in jets~\citep{2011NewA...16...46B}. 
Assuming the separation in the fluid comoving frame as $\bar{l}$, 
the reconnection timescale can be written as: $\bar{\tau}_{\rm dissip} = \bar{l}/v_{\rm in}$ in the fluid comoving frame. 
If we assume the reversing of the magnetic field direction occurs every one-Kepler rotation time at a radius $r$,  
the timescale in the blackhole rest frame can be written as, $\tau_{\rm rot} \sim \pi r_M (2r/r_M)^{3/2}/ c$ where $r_M$ is the Schwarzshild radius. 
Hence, $\bar{l}$ can be written as, $\Gamma_{\rm jet} c \tau_{\rm rot}$. 
Since the dissipation timescale in the central blackhole comoving frame, $\Gamma_{\rm jet} \bar{\tau}_{\rm dissip}$, should be less than 
the jet propagation time, $l_{\rm jet}/c$, we obtain
\begin{align}
  \label{eq:5.3}
  \frac{v_{\rm in}}{c} &\gtrsim \frac{\bar{l}}{l_{\rm jet}} \Gamma_{\rm jet} 
  \sim \pi \Gamma_{\rm jet}^2 \left( \frac{r_M}{l_{\rm jet}} \right) \left( 2 \frac{r}{r_M}\right)^{3/2}
  \nonumber
  \\
  &\sim
  \begin{cases}
  &1.9 \times 10^{-3} \left(\frac{l_{\rm jet}}{60[\textrm{pc}]}\right)^{-1} \left(\frac{r}{3 r_M}\right)^{3/2} \left(\frac{r_M}{10^{-4}[\mathrm{pc}]}\right) \left(\frac{\Gamma_{\rm jet}}{5}\right)^2 
    \quad \textrm{(radio)},
  \\
  &3.3 \times 10^{-4} \left(\frac{l_{\rm jet}}{0.5[\textrm{kpc}]}\right)^{-1} \left(\frac{r}{3 r_M}\right)^{3/2} \left(\frac{r_M}{10^{-4}[\mathrm{pc}]}\right) \left(\frac{\Gamma_{\rm jet}}{6}\right)^2
    \quad \textrm{(HST)},
  \end{cases}
\end{align}
where the upper and lower values are based on the radio and HST data of M87~\citep{2012MPLA...2730030R}. 
Here we assume the reversing of the magnetic field direction occurs at the innermost secure radius $r = 3 r_M$.  
This indicates the both observation results can be explained by turbulent reconnection. 
\\\\
\textit{--Gamma Ray Bursts}\\
It is suggested that GRB can be explained by Poynting-dominated plasma model, 
and \citet{2011ApJ...726...90Z} suggested a model called ICMART model that can explain GRB including prompt emission spectral curves~\citep{2014ApJ...782...92Z}. 
In the model, the reconnection rate is assumed a relativistic value, the minimal value is around $0.1c$. 
As indicated in Figure 2, this value of reconnection rate will be obtained by turbulent reconnection in Poynting dominated plasma. 
The model also discussed a possible role of reconnection outflow assuming Alfv\'enic velocity, $\gamma_A \sim \sqrt{1+\sigma}$. 
In our calculations, a relativistic outflow was observed only locally and intermittently; 
the averaged outflow velocity was a sub-relativistic velocity up to $\sim 0.3c$ even in the high-$\sigma$ regime, 
which is commonly seen in the relativistic MHD reconnection resulted from the tearing instability. 
If we see a smaller scale, the collisionless plasma regime will appear in which the outflow velocity is Alfv\'en velocity in the high-$\sigma$ regime~
\citep{2015PhRvL.114i5002L}.

Recently, \citet{LM2015} proposed a new scenario for the gamma-ray bursts powered by turbulent reconnection 
based on kink instability of relativistically magnetized jets~\citep{2012ApJ...757...16M}. 
In this model, 
the authors considered a turbulence induced by the kink instability, 
and applied it to the turbulent reconnection model. 
They showed that their model can provide a good fit to the dynamics of GRBs. 
Note that this model may also be able to be applied to AGN jets, 
which may result in a different condition from Equations (\ref{eq:5.3}). 
\\\\
\textit{--General Remarks}\\
Finally, we give several comments which can be applied to any phenomena in general if turbulent reconnection works. 
First, our results, such as the Figure 2, indicates that the maximum reconnection rate is obtained when $v_{\rm inj}/c_A \sim 0.2$. 
Assuming the Alfv\'enic mode, $\delta B / B_0 \sim \delta v_{\perp}/c_A$, where $B_0$ and $\delta B$ are the background and fluctuation of magnetic field, respectively, 
$\delta v_{\perp}$ is the fluctuation velocity perpendicular to $B_0$, 
we can expect that the turbulent reconnection is efficient 
if an observed magnetic fluctuation of some phenomena is around $\delta B \sim 0.2 B_0$. 
In addition, we can expect that such a low magnetic fluctuation will allow a high polarization degree of synchrotron radiation 
observed by \citet{2003Natur.423..415C}. 
Second, 
it is well-known that magnetic reconnetion is also related to particle acceleration. 
For example, many Particle-In-Cell simulations indicates that 
the energy spectrum index obtained in relativistic collisionless pair-plasma reconnection is around $-1.5$
~\citep{2001ApJ...562L..63Z,2012ApJ...750..129B,2014ApJ...783L..21S,2015ApJ...806..167G}. 
On the other hand, \citet{2005AA...441..845D} found that 
the turbulent reconnection is also a location of the first-order Fermi process, 
and they obtained a little softer energy spectrum index, $-2.5$. 
Non-thermal particle energy index is directly related to the observed synchrotron spectrum, 
and the above energy spectrum index may be useful tool to determine the location of turbulent reconnection. 
Applications of turbulent reconnection model to some AGNs are provided 
in \citep{2015MNRAS.449...34K,2015ApJ...802..113K,2015ApJ...799L..20S}. 
More comprehensive discussion of recent observations is given in~\citet{2015SSR}. 

\section{\label{sec:sec6}Discussion And Conclusion}

In this paper, we investigated turbulent reconnection in relativistic plasmas from the matter dominated to Poynting dominated cases 
using the relativistic resistive MHD model. 
The results show that 
the turbulence can enhance magnetic reconnection even in relativistic plasmas, 
and can be a candidate for a fast reconnection process. 
We found 
the reconnection rate in turbulence shows the following 3 characteristic phase depending on the velocity of the injected turbulence: 
(1) LV99 region (incompressible turbulence); (2) saturation region giving maximum rate; (3) reducing due to the compressibility. 
The saturation occurs when the compressible component become dominant, typically around $v_{\rm c}^2 /v_{\rm i}^2 \gtrsim 0.4$ 
at which the maximum reconnection rate is about $0.05$. 
This shows that the LV99 expressions for incompressible fluid should be modified to account for compressibility 
as we have done in Equations (\ref{eq:4.3}) and (\ref{eq:4.4}). 
Interestingly, \citet{2013PhRvE..87a3019B} showed that 
dilatation of fluid, $\nabla \cdot {\vec v} > 0$, reduces the effective energy cascade rate $\epsilon_{\rm eff}$. 
All of our numerical results showed dilatation, and the reconnection rates are indeed reduced. 
This indicates that the turbulent reconnection rate may become larger than LV99's prediction if compression of the turbulence occurs, 
such as the MHD turbulence driven by collisions of magnetized blobs~\citep{2011ApJ...734...77I,2015ApJ...805..163D}. 

Finding a fast reconnection process is one of the most important topics in plasma physics, 
and a considerable number of studies have been conducted on it for a long time. 
Turbulence is a very general process in high-Reynolds number plasmas, 
so that turbulent reconnection can appear in many kinds of phenomena, such as astrophysical phenomena, nuclear fusion, and laser plasma. 
In particular, our work investigated the extension of turbulent reconnection to relativistic plasma with compressible turbulence, 
which allows us to apply this process to many high energy astrophysical phenomena, such as flares in pulsar wind nebulae, gamma ray bursts, and relativistic jets. 


\acknowledgments
We would like to thank John Kirk, S\'ebastien Galtier, Supratik Banerjee 
for many fruitful comments and discussions. 
We also would like to thank our anonymous referee for a lot of fruitful comments on our paper. 
Numerical computations were carried out on the Cray XC30 
at Center for Computational Astrophysics, CfCA, of National Astronomical Observatory of Japan.
Calculations were also carried out on SR16000 at YITP in Kyoto University. 
This work is supported in part by the Postdoctoral Fellowships for Research Abroad program by the Japan Society for the Promotion of Science No. 20130253 
and also by the Research Fellowship for Young Scientists (PD)
by the Japan Society for the Promotion of Science 
No.~20156571 (M.T.). 
One of the author (A. Lazarian) is supported by NSF AST 1212096. 
This work is also supported in part by Grants-in-Aid for Scientific Research 
from the MEXT of Japan, 15K05039 (T.I.). 







\end{document}